\documentclass[11pt,twoside]{article}
\usepackage{color}

\usepackage{asp2006}
\usepackage{epsf}
\usepackage{psfig}
\usepackage{lscape}

\begin{document}
\title{21-cm absorbers at intermediate redshifts.}   
\author{N. Gupta$^1$, R. Srianand$^2$, P. Petitjean$^3$, P. Noterdaeme$^2$, D. J. Saikia$^4$}
\affil{ $^1$Australia Telescope National Facility, CSIRO, Australia\\ 
$^2$IUCAA, India \\
$^3$Institut d'Astrophysique de Paris-CNRS, France \\
$^4$NCRA-TIFR, India}

\begin{abstract} 
Damped Lyman-$\alpha$ systems (DLAs) seen in the spectra of high-$z$ QSOs allow us to probe
the physical conditions in protogalaxies.  Our understanding of physical conditions in
DLAs at high-$z$ is primarily based on the absorption lines of H$_2$ molecules and fine-structure
transitions.  Another important way of probing the thermal state of interstellar medium in these
systems is by studying the 21-cm absorption in the spectra of background quasars.
Here we report the main results of our GMRT survey to search for
21-cm absorption in a representative and unbiased sample of 35 DLA candidates at 1.10$\le$$z$$\le$1.45.
Our sample of DLA candidates is drawn from the strong
Mg~{\sc ii} systems in SDSS DR5 and has resulted in discovery of 9 new 21-cm
absorbers.  Prior to our survey only one 21-cm absorber was known in the
redshift range: 0.7$\le$$z$$\le$2.
This survey has allowed us to investigate the dependence of detectability of 21-cm absorption on 
the properties of UV absorption lines detected in SDSS spectra and estimate the number per unit 
redshift of 21-cm absorbers.
Our GMRT survey provides a representative sample of systems that
can be used in combination with various follow-up observations: (1) for investigating the
physical conditions in the absorbing gas using spin temperature (T$_{\rm S}$) measurements,
(2) for investigating the effect of metallicity and dust content on the detectability
of 21-cm absorption, (3) for
studying the morphology of the absorbing gas and (4) for probing the time evolution of
various fundamental constants.
Results from the first phase of our survey are presented in Gupta et al. (2007). Detailed description of the 
entire sample and results from the survey are presented in Gupta et al. (2009).

\end{abstract}


\section{Introduction}   
Observations of high-$z$ galaxies suggest that
the global comoving star-formation rate density
peaks at $1\le z\le 2$ and then sharply decreases
towards $z\sim0$ (e.g. Madau et al. 1996, Hopkins 2004).
The determination of the mass density of the gas and its content (molecules, dust and cold H~{\sc i} gas)
over the same redshift range provides an independent and complementary understanding of the redshift
evolution of star-formation at similar epochs.
While the H~{\sc i} content of galaxies can be best probed
by the surveys of 21-cm emission, limited sensitivity
of current radio telescopes does not allow them to reach beyond the
local Universe (e.g., Zwaan et al. 2005). On the contrary, detection
of H~{\sc i} in the spectra of distant QSOs in the form of
damped Lyman-$\alpha$ absorption provides a luminosity unbiased way
of probing the evolution of the H~{\sc i} content in the universe
(Wolfe, Gawiser \& Prochaska 2005).

Our understanding of physical conditions in DLAs at $z\ge2$ is
largely based on the analysis of H$_2$ and/or atomic fine-structure transitions (Ledoux, Petitjean \& Srianand 2003).
Unfortunately for the time being
the above mentioned tracers can not be used to probe the
physical state of the absorbing gas at $z\le1.8$ because the
useful transitions are located below the atmospheric cut-off.
It has been shown by Rao, Turnskek \& Nestor (2006) that DLAs essentially
have Mg~{\sc ii} rest equivalent width, W$_{\rm r}$(Mg~{\sc ii}$\lambda2796$)$\ge$0.6\AA.
Therefore, the search of 21-cm absorption in a sample of strong Mg~{\sc ii} absorbers
is an unique way to probe the redshift evolution of physical conditions in DLAs
like absorption systems at intermediate and low-$z$.

 \begin{figure}
 \plottwo{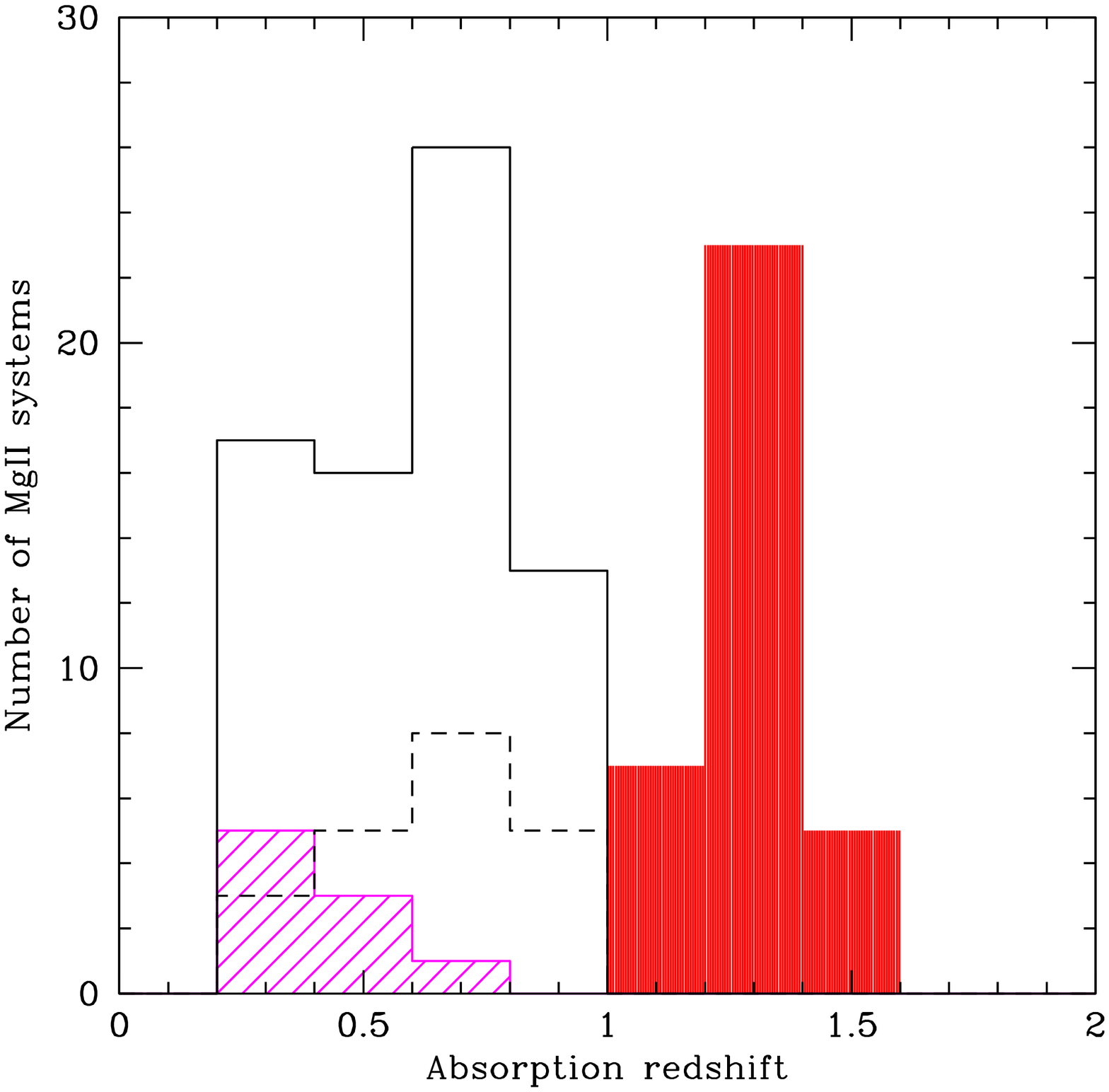}{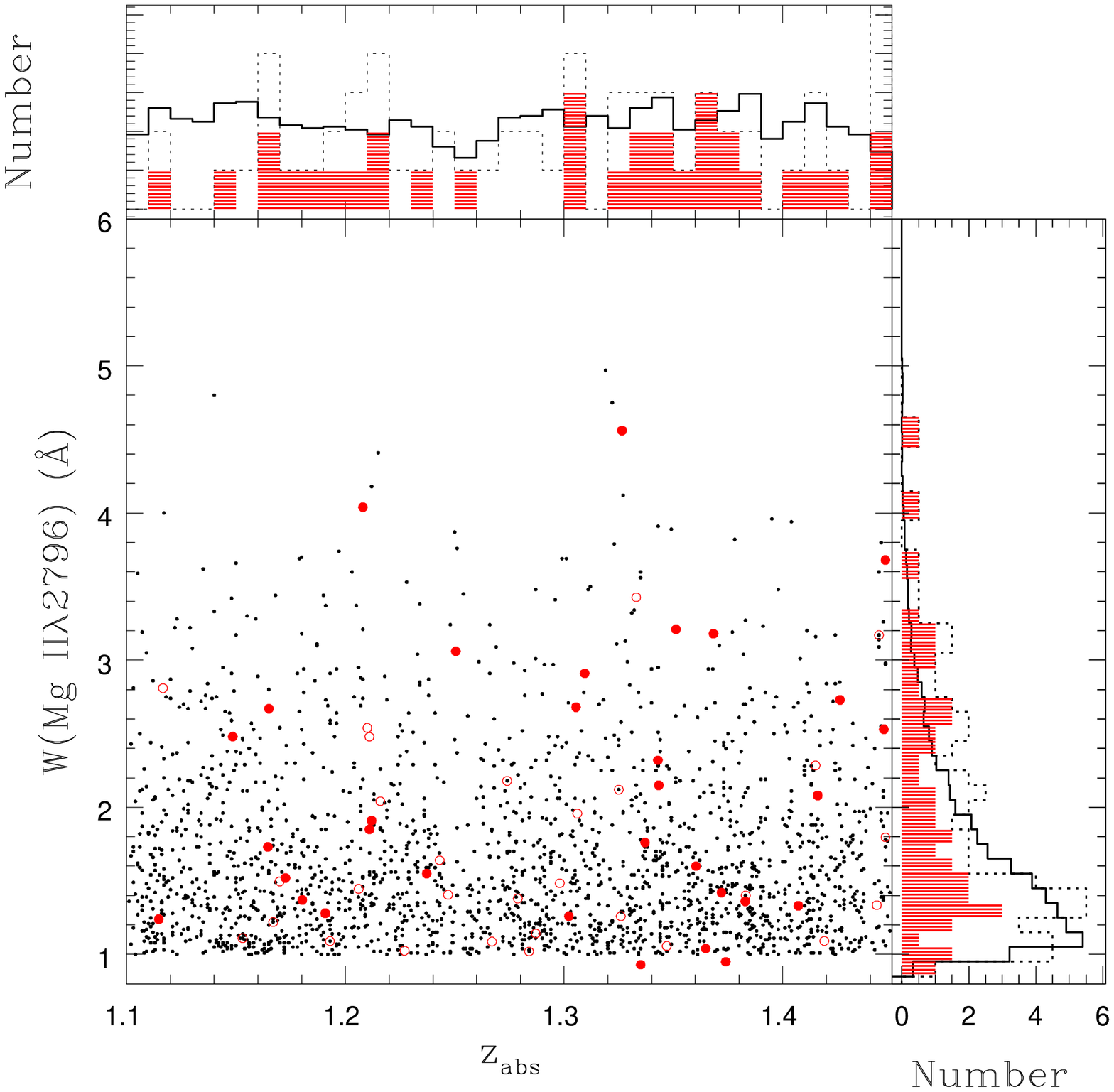}
 \caption{{\itshape Left:\/} Redshift distribution of Mg~{\sc ii} systems that were
searched for 21-cm absorption. The filled histogram is the GMRT
sample of 35 Mg~{\sc ii} systems presented in this paper (33 of these absorption
systems have W$_{\rm r}$(Mg~{\sc ii}$\lambda$2796)$\ge1${\AA}).
The solid line histogram is for the sample of Lane (2000).
The hatched histogram corresponds to 21-cm detections in this sample.
The distribution for the W$_{\rm r}$(Mg~{\sc ii}$\lambda$2796)$\ge 1${\AA}
sub-set of these systems is given by the dashed line histogram.
 {\itshape Right:\/} Mg~{\sc ii}$\lambda$2796 rest equivalent width
as a function of the absorption redshift. Small dots are for the whole sample of
Mg~{\sc ii} systems. Open and filled circles are for the complete sample
of Mg~{\sc ii} systems in front of radio-loud QSOs with flux density greater than 50 mJy.
The filled circles are the ones observed in this survey.
Histograms for the distribution of $z_{\rm abs}$ and W$_{\rm r}$ are also shown.
The solid, dashed and shaded histograms are for the overall Mg~{\sc ii} sample,
the Mg~{\sc ii} sample in front of radio-loud QSOs and the sample
observed with GMRT respectively.
}
 \end{figure}


%
%
\section{Our GMRT sample}
Our sample is drawn from the identification of strong Mg~{\sc ii} systems, 
W$_{\rm r}$(Mg~{\sc ii} $\lambda2796$)$\ge$1.0\AA, by 
Prochter, Prochaska \& Burles (2006, hereafter P06) in SDSS DR3 and by us using our 
automatic procedure for additional systems in DR5.  
We select the absorbers that are in the redshift range: 1.10$\le z_{abs}\le$1.45 such that 
the redshifted 21-cm frequency lies in the GMRT 610-MHz band.  
GMRT is the only radio telescope available at present in the {\it relatively} RFI-clean
environment (say compared to Green Bank Telescope or Westerbork Synthesis Radio Telescope)
for covering this redshift range.
These absorbers are then cross-correlated with NVSS and FIRST surveys to select the Mg~{\sc ii} 
absorbers in front of compact radio sources brighter than 50\,mJy and hence suitable 
for the 21-cm absorption search.
We plot in Fig.~1 the redshift distribution of the 35 Mg~{\sc ii} absorption systems
observed as part of our GMRT survey along with the sample of Lane (2000). Latter is the only 
large survey at low-$z$ for which both detections
and non-detections are systematically reported.
It includes 62 systems observed with the WSRT and 10 other
systems from the literature satisfying their selection criterion (see Lane 2000 for details).
The detections shown as a hatched histogram include
detections reported in Lane (2000) together with detections
from better quality data by Kanekar \& Chengalur (2003) and
Curran et al. (2007) for systems that were originally reported as
non-detections. In the same figure, the filled histogram shows the distribution of
Mg~{\sc ii} systems in our GMRT sample.
For equivalent width cutoff of $\sim$ 1{\AA}, our GMRT sample
has more than twice the number of systems investigated by Lane (2000).

\begin{figure}
\centerline{{
\psfig{figure=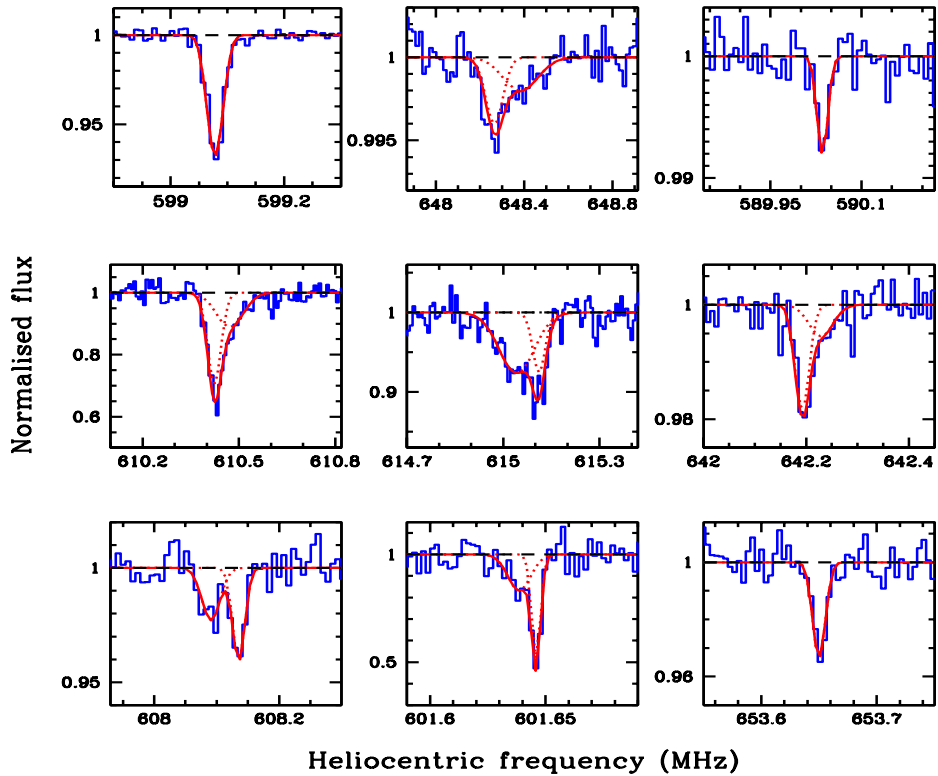,height=11.0cm,width=13.0cm,angle=0}
}}
\caption[]{GMRT spectra of detected 21-cm absorption lines. Individual Gaussian components and
resultant fits to the absorption profiles are overplotted as dotted and continuous lines respectively.
}
\label{mg2det}
\end{figure}

We observed 35 Mg~{\sc ii} systems
with GMRT 610-MHz band using in total $\sim$400\,hrs of telescope time mostly
spread over the years 2006-2008.
For our survey, we have usually used the 1\,MHz baseband bandwidth split into 128 frequency
channels yielding a spectral resolution of $\sim$4\,km\,s$^{-1}$.
GMRT data were reduced using the NRAO Astronomical Image Processing System
(AIPS) following the standard procedures.

\section{Results and Discussion}
We present the results of a systematic Giant Metrewave Radio Telescope
(GMRT) survey of 21-cm absorption in a representative and
unbiased sample of 35 strong Mg~{\sc ii} systems in the redshift range:
$z_{abs}\sim$1.10$-$1.45, 33 of which have W$_{\rm r}\ge$1\,\AA.
The survey using $\sim$400\,hrs of telescope time has resulted in
9 new 21-cm detections and good upper limits for the remaining 26 systems (Fig.~\ref{mg2det}).
This is by far the largest number of systems detected in a single systematic
survey in a narrow redshift range. Two of these systems also show 2175~{\AA} dust feature
at the redshift of the absorbers (Srianand et al. 2008).
Results from the first phase of our survey are presented in Gupta et al. (2007).
Detailed description of the entire sample and results from the survey are presented in 
Gupta et al. (2009).  In the following we summarise the main results. 
%

 \begin{figure}
 \plottwo{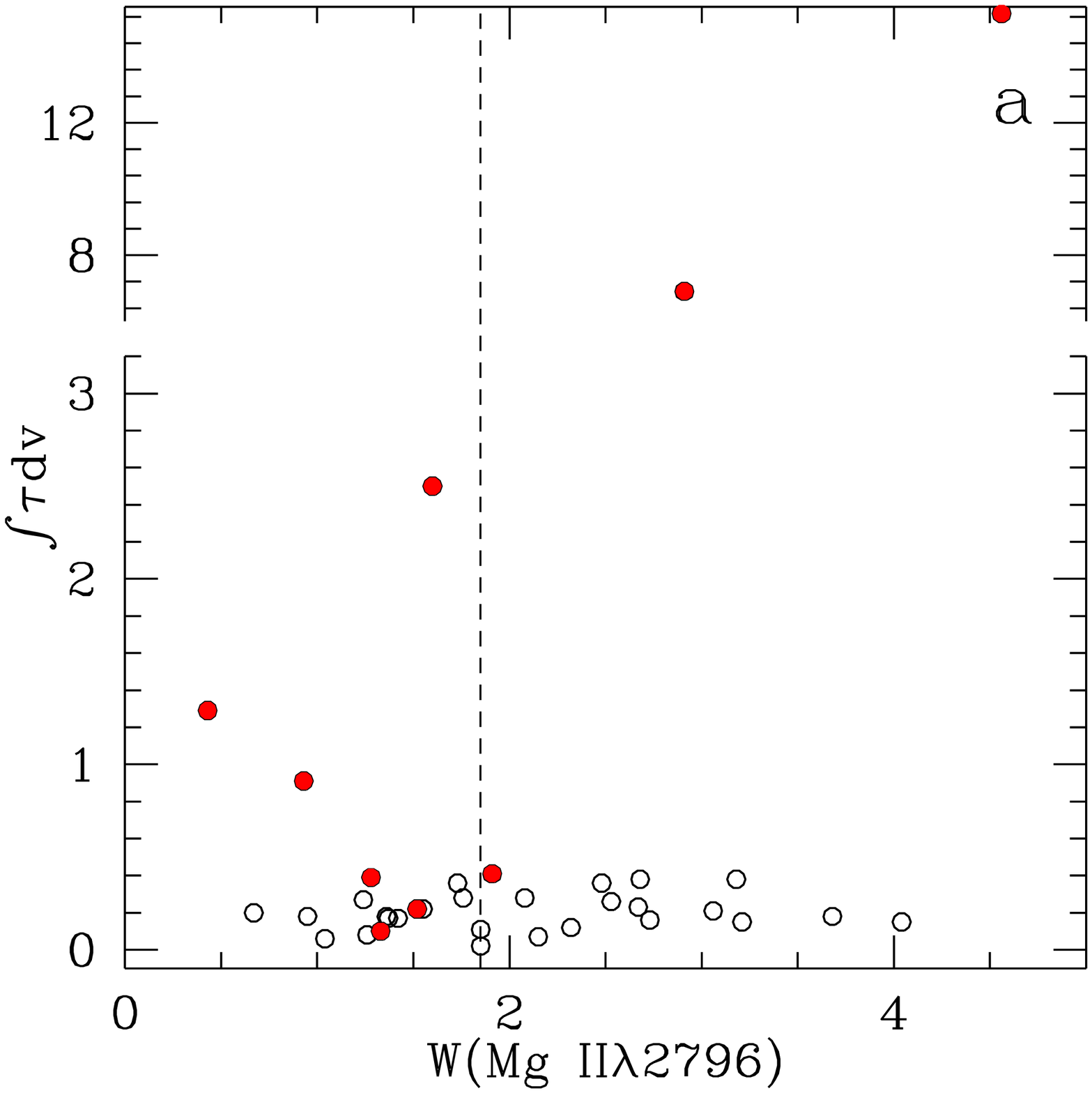}{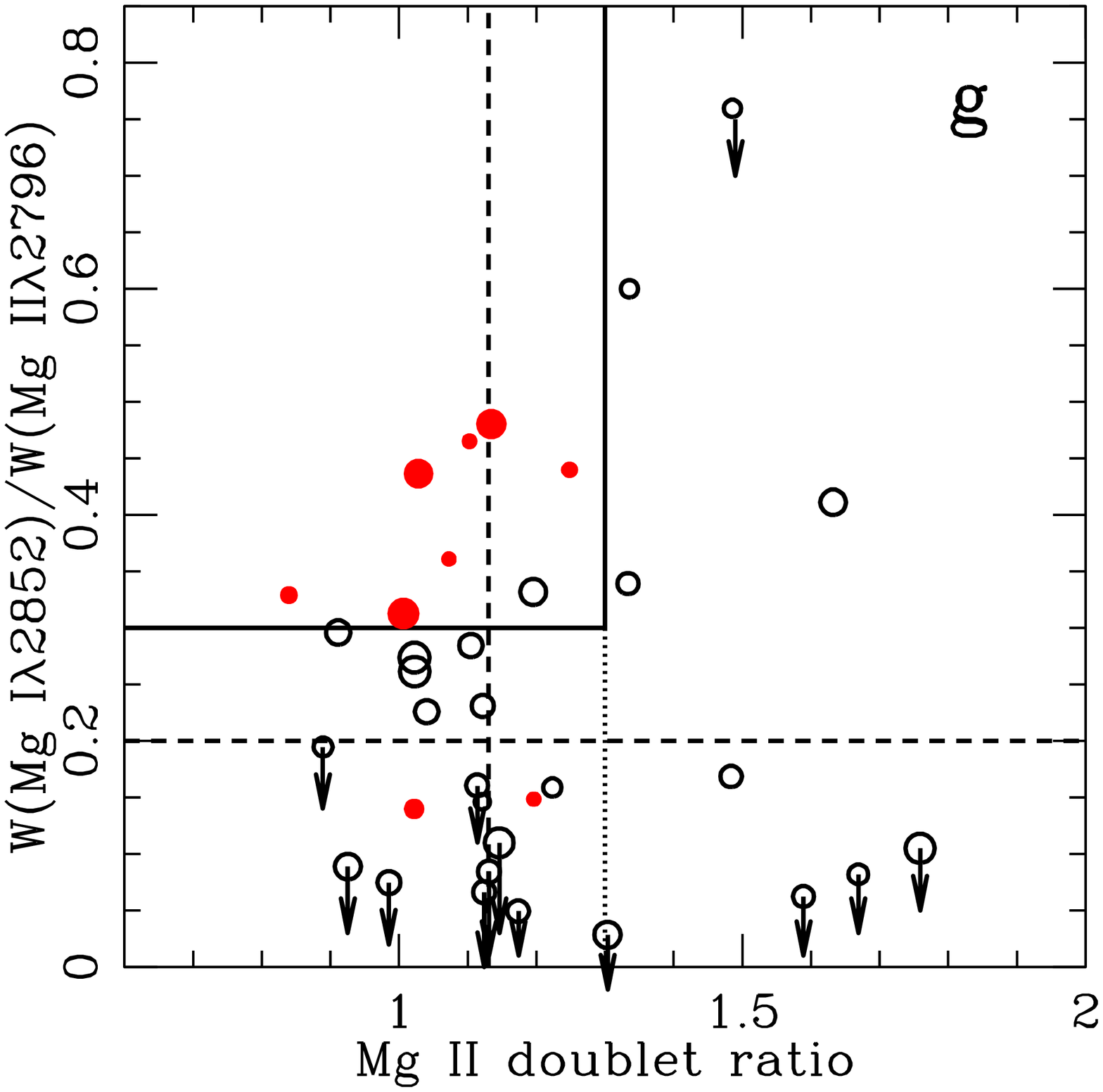}
 \caption{{\itshape Left:\/} 
 The integrated 21-cm optical depth is plotted against W$_{\rm r}$(Mg~{\sc ii}$\lambda$2796). 
In the case of detection (filled circle) the optical depth is obtained
by integrating over the observed absorption profile. Open circles are
for 21-cm non-detections in which case limits are obtained by integrating the optical depth 
over a Gaussian component with peak optical depth corresponding to the 3$\sigma$ rms limit in
the continuum and width 10 km~s$^{-1}$.
 {\itshape Right:\/} 
The ratio W$_{\rm r}$(Mg~{\sc i})/W$_{\rm r}$(Mg~{\sc ii}) is plotted against the Mg~{\sc ii} doublet ratio. 
In both the panels,  the vertical dashed line gives the
median value. Other lines indicate limiting values and/or allowed ranges as discussed in the text.
}
 \vskip -14.0cm
 \begin{picture}(400,400)(0,0)    
 \color{white} 
 \put(160,260){\circle*{15}}
 \put(340,260){\circle*{15}}
 \end{picture}

 \label{uv21}
 \end{figure}


We study the dependence of detectability of 21-cm absorption on different
properties of the UV absorption lines detected in the SDSS spectra (Fig.~\ref{uv21}).
We find that if absorption systems are selected with a Mg~{\sc ii} doublet ratio, DR~$<$~1.3, and
a ratio W$_{\rm r}$(Mg~{\sc i})/W$_{\rm r}$(Mg~{\sc ii})~$>$~0.3, the success rate
for 21-cm detection is very high (up to $90$\%; see {\it right} panel of Fig.~\ref{uv21}).
We notice that the detections found in a low-$z$ sample by Lane (2000) also obey
these joint constraints (see left panel of Fig.~\ref{uv21}). 
In our sample, we find an apparent paucity of 21-cm absorption among
systems with W$_{\rm r}$(Mg~{\sc ii}$\lambda$2796)$>$1.8 {\AA},
the median W$_{\rm r}$ of our sample. This is contrary to what has been seen at
low-$z$ (Lane 2000).
Interestingly most of these high W$_{\rm r}$ systems
have high DR and low values of W$_{\rm r}$(Mg~{\sc i})/W$_{\rm r}$(Mg~{\sc ii}).
This strongly suggests that the equivalent width in these systems is dominated
by velocity spread and not by line saturation.

We estimate the number of 21-cm absorption systems per unit redshift interval
for a given limiting value of the integrated 21-cm optical depth
and W$_{\rm r}$(Mg~{\sc ii}$\lambda$2796).
We show that 
the fraction of Mg~{\sc ii} systems with 21-cm absorption and the
number per unit redshift decrease from $z\sim0.5$ to $z\sim1.3$.
The decrease is larger when we use higher equivalent width cutoff.
Using a sub-sample of compact sources, with high frequency VLBA observations
available, we show that this can not be accounted for by simple covering factor effects.
As mentioned above and based on the available data, it appears that most likely
the main reason behind this cosmological evolution is the decrease of the CNM covering factor
(and volume filling factor) in the strong Mg~{\sc ii} absorbers.
Indeed, it is known that the number of Mg~{\sc ii} systems per unit redshift increases
with increasing redshift. The evolution is steeper for stronger systems
(Steidel \& Sargent, 1992 and P06 for recent reference).
Using the data of Steidel \& Sargent (1992), Srianand (1996)
found that the strongest redshift evolution was seen among the
Mg~{\sc ii} absorbers with W$_{\rm r}$(Fe~{\sc ii})$\lambda$2383/W$_{\rm r}$(Mg~{\sc ii})$\lambda$2796$<$0.5.
This clearly means the physical conditions in strong Mg~{\sc ii} absorbers are
different at high and low-$z$.

We have estimated the velocity spread of the 21-cm absorption systems using the
apparent optical depth method (Ledoux et al. 2006). 
We do not find any statistically significant correlation between
W$_{\rm r}$(Mg~{\sc ii}$\lambda$2796) and the 21-cm velocity width in our sample.
A marginal correlation is found for the low-$z$ sample.
The absence of correlation in the high-$z$ sample is related to
the lack of 21-cm absorbers among Mg~{\sc ii} systems with 
W$_{\rm r}$(Mg~{\sc ii})$>$1.8 {\AA} in the GMRT sample. This
is probably due to a true evolution with redshift of the physical state of the
Mg~{\sc ii} systems and consistent with the idea that
the Mg~{\sc ii} equivalent width is mostly correlated with the overall kinematics of the
gas in the absorbing system and not with the column density in the
component associated with the cold gas.
When high spectral resolution data are available, we note that the 21-cm absorption
is not always associated with the strongest Mg~{\sc ii} component.

%
We selected systems with W$_{\rm r} \ge$1\,\AA~ but detected by chance a 21-cm absorption
in a system with  W$_{\rm r}$~=~0.43\,\AA (at $z_{\rm abs}$~=~1.3710 toward J0108$-$0037).
Efforts are underway at GMRT to extend our survey to weaker (W$_{\rm r} \le$1\,\AA~) Mg~{\sc ii} systems.
This will be crucial for understanding the physical state of Mg~{\sc ii} systems and
to determine the detectability of 21-cm absorption versus W$_{\rm r}$.
Ideally one would like to estimate the number density of 21-cm absorbers and measure the cosmological
evolution without preselection from the UV absorption lines. This can be achieved only by a blind
survey of 21-cm absorption in front of radio loud QSOs.
It will be possible to embark upon
such a survey with the upcoming Square Kilometer Array (SKA) pathfinders such as the 
Australian Square Kilometer Array Pathfinder (ASKAP) and Karoo Array Telescope (meerKAT).
In particular, the ASKAP with its  instantaneous wide bandwidth
of 300\,MHz and large field of view (30\,degree$^2$) is an ideal instrument for this
(Johnston et al. 2008).
An ASKAP survey with 150 pointings of 16~hrs each (i.e 2400 hrs in total) in the 700-1000 MHz frequency
band would yield detection of $\sim$100 to 250 intervening 21-cm absorbers in the redshift range 0.4$\le z \le$1.
%

As the energy of the 21-cm transition is proportional to $x=\alpha^2 G_p/ \mu$,
high resolution optical and 21-cm spectra can be used together to probe the combined
cosmological variation of these constants (Tubbs \& Wolfe 1980).
Our GMRT survey provides systems in a narrow redshift range in which this
measurement can be done.
Thus high resolution optical spectroscopy of the corresponding QSOs are suitable
to perform this test at $z\sim1.3$.

\acknowledgements 
We thank Rajaram Nityananda for useful discussions and
DDTs, and the GMRT staff for
their co-operation during the observations.
GMRT is run by the National Centre for Radio Astrophysics of the Tata Institute of 
Fundamental Research.  
We acknowledge the use of SDSS spectra from the archive (http://www.sdss.org/) and the 
radio images from NVSS and FIRST surveys.


\end{document}